\documentclass{article}

\usepackage{graphicx}
\usepackage{cite}
\usepackage[cmex10]{amsmath}

\begin{document}
\sloppy
\title{\bf\LARGE Information Nano-Technologies:
Transition from Classical to Quantum}

\author{Alexander Yu.\ Vlasov}

\maketitle

\begin{abstract}
In this presentation are discussed some problems, relevant with
application of information technologies in nano-scale systems
and devices. Some methods already developed in quantum information 
technologies may be very useful here. Here are considered two
illustrative models: representation of data by quantum bits and 
transfer of signals in quantum wires.\\
{\bf keywords--quantum; information; nanotechnologies}
\end{abstract}

\section{Introduction}
\label{Sec:Intro}

Let us recollect well known Feynman's talks, relevant to presented theme. 
The first one, is the Caltech lecture ``There's plenty of room at the bottom'' 
in 1959 \cite{plenty} often is considered between the origins of nanotecnnologies. 
The second one, is the keynote speech ``Simulating physics with computers'' \cite{feysim} 
in 1981 at the conference PhysComp'81 ``Physics and Computations'' about physical
background of computing and information technologies. Main part of 
this talk was devoted to quantum processes. 

In this speech was established some ideas, essential for the development of 
quantum computations and communication, but not only that.
The simulation --- is detailed modeling of a physical process. 
For quantum systems it is the especial challenge, because the formulations 
of the quantum theory is often similar with a ``black box'' \cite{ngnr,vlnl} description.

A positive result of the research of {\em physics of computations} 
was understanding of principle possibility of information processing by devices 
with elements of the atomic size. Sometimes it was even necessary to critically 
revisit some widespread ideas. For example, elements in such a scale often may be 
more adopted for {\em reversible} operations, but most gates in standard computer  
design are {\em irreversible}.

Charles Bennett suggested a model of a reversible Turing 
machine and even denoted a similarity of such a model with DNA and RNA \cite{ben73}. 
The reversible Turing machine has direct generalization on quantum systems and it was 
demonstrated in few works of Paul Benioff, including the presentation on already 
mentioned PhysComp'81 \cite{ben82}. Feynman's representation is more close to 
the modern description of computers by gates and circuits, but uses specific 
attributes of quantum mechanics \cite{feysim}.

\smallskip

In Section \ref{Sec:qubits} is reminded an abstract quantum analogue of
classical bit. In Section \ref{Sec:quprop} are briefly discussed distribution
of signals in nanosystems and relevant quantum effects. In Section \ref{Sec:perf}
is revisited so-called perfect state transfer. Section \ref{Sec:Weyl}
is devoted to Weyl commutation relations. 

\section{Quantum bits}
\label{Sec:qubits}

 There is widespread notation $|0\rangle$ and $|1\rangle$ for two basic states of a 
quantum system, which often is called {\em quantum bit} or {\em ``qubit''}. Feynman 
had used for manipulations with a qubit expressions with formal operators of 
{\em annihilation} and {\em creation}: ${\bf a}|1\rangle = |0\rangle$,
${\bf a}^*|0\rangle = |1\rangle$,
${\bf a}\,{\bf a}^* + {\bf a}^* {\bf a} = {\bf 1}$, 
${\bf a}^2 =({\bf a}^*)^2 = {\bf 0}$.

Let us consider a set with eight qubits. Basic states of such a ``quantum byte'' 
may be described as strings of zeros and units: $|00000000\rangle$, $|00000001\rangle$,
$\ldots$, $|11111111\rangle$. It can be simply estimated, there are $2^8 = 256$ 
basic states or $2^N$ for a system with $N$ particles. It is in agreement rather with 
the principles of quantum mechanics, than with the classical case. In ``computer 
notation'' it is clear enough even without more pedantic consideration of tensor 
product of linear spaces describing a state of the quantum system.

An illustrative classical picture still exists for one qubit and any state 
may be represented by direction of some ``arrow,'' like two basic states: 
``spin up'' and ``spin down''. 
For a classical case description of such a system also demands two parameters, 
{\em e.g.}, the Euler angles. But this visual correspondence 
disappears in a case with few systems, because in the classical world for description 
of $N$ ``arrows'' it would be necessary to use only $2N$ parameters instead of $2^N$.

\begin{figure}[!t]
\centering
a)~
\includegraphics{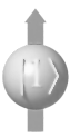}~
\includegraphics{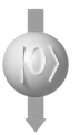}~
\includegraphics{spin1ket.eps}~
\includegraphics{spin0ket.eps}~
\includegraphics{spin1ket.eps}~
\includegraphics{spin0ket.eps}~
\includegraphics{spin0ket.eps}~
\includegraphics{spin1ket.eps}

b)~
\includegraphics{spin1ket.eps}~
\includegraphics{spin0ket.eps}~
\includegraphics{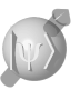}~
\includegraphics{spin0ket.eps}~
\includegraphics{spin1ket.eps}~
\includegraphics{spin0ket.eps}~
\includegraphics{spin0ket.eps}~
\includegraphics{spin1ket.eps}

\caption{Quantum spin chains}
\label{fig_chain}
\end{figure}

Of course, classical bits may be represented in the classical model,
as a discrete set with $2^N$ elements inside of a space with $2N$ 
{\em continuous} parameters. The quantum model with $2^N$ parameters 
also includes this set (Figure \ref{fig_chain}a), and
here each element {\em directly} conforms to a continuous parameter. 

The quantum model corresponds to a classical one, if only states of {\em separate}
qubits may differ from two fixed options of usual bit (Figure \ref{fig_chain}b). 
The difference $2^N\!\! - 2N$ is an approximate estimation of {\em ``non-classicality''} 
and it grows very fast with the number of systems $N$. 

{\em This consideration of complexity has relevance with presented theme,
because the nano-scale domain describes aggregates with more than one quantum 
system, but it is still not big enough to use statistical laws. The quantum 
theory of information provides the convenient language for description of systems 
with not very big amount of elements due to appropriate level of abstraction.
\samepage

}

\begin{figure}[!t]
\centering
\includegraphics{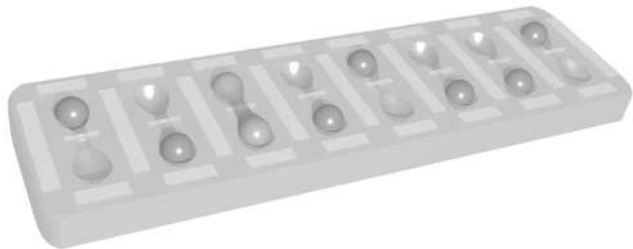}
\caption{Quantum (double) dots}
\label{fig_ddots}
\end{figure}

For example, the same model may be applied to different quantum systems.
A spin one-half system was used in the visual picture above, but the {\em qubit}
is a model for many other systems with two states, like photons or
quantum dots. Multi-qubit systems like ``quantum byte'' also may be associated 
not only with spin chain (Figure \ref{fig_chain}), but with quantum dots arrays 
(Figure \ref{fig_ddots}) and other implementations \cite{qcroad}.

\section{Quantum signal propagation in nanosystems}
\label{Sec:quprop}

Let us discuss now application of some ideas to next generations of nanotechnologies. 
Such devices are still in a state of development and it may be reasonable to pay 
attention to processes in biosystems. Recent time active research is carried out 
with respect to descendants of most ancient ``nanodevices'' existing on the Earth about 
three milliards years or so. 

It is the light-harvesting complex of some microorganisms. 
The importance of quantum effects for this case is already almost impossible to deny. 
The significant contribution for understanding here is due to works with participation 
of experts in quantum theory of information \cite{jo08, ll08, pl09, wh09}.

Let us consider a problem of the effective transfer of 
absorbed photon energy to different elements of a nanosystem. In the biological 
systems mentioned above the effectiveness may be about 99\%. It is astonishing 
with taking into consideration of quantum uncertainty, because it apparently 
should hinder the optimal transfer.

Yet biophysical processes in such systems formally look as not relevant with 
information transfer, a set of problems and methods applied there are very similar 
with the statement of a question about an effective transmission of signals in 
a nanodevice with taking into account of quantum effects. {\em E.g.}, in paper 
\cite{ll08} is suggested a transfer model based on a quantum analogue of the 
random walk. In the classical case a chaotic motion may be considered as a
quite effective way of a transport in complicated compact systems.

Uncertainty of positions and trajectories in quantum mechanics needs for the special 
consideration. There is well known approach with the suggestion about state localization 
due to the interaction with the environment. It is one possible explanation
of the transition from quantum to classical world \cite{zur03}. Related ideas about 
decoherence assisted transport may be more or less directly used in some models about 
effective energy transfer in the light-harvesting complexes \cite{ll08,pl09}.

Some classical models are tested very well for the macroscopic level and it is not 
clear at that scale they are still work for nanosystems. It is reasonable to check 
a possibility of description of the effective transfer without appealing to 
semi-empiric regularities acting on the boundary between quantum and classical 
world \cite{zur03}. 

Indeed, the possibility of the ``perfect'' transfer of an excitation in ``purely'' 
quantum approach was also found recently \cite{ek04}. Similar methods was only briefly
mentioned in the relation with the biophysical systems discussed above \cite{wh09}.

In the works \cite{vl07,vl08} were also considered some aspects of this approach,
appropriate to the present discussion. 
It can be said, the model of perfect transfer is an analogue of shift register: 
$10000000$ $\rightarrow$ $01000000$ $\rightarrow$ $\ldots$ $\rightarrow$ $00000001$. 
Here unit corresponds to the excited state. Coefficients describing strength of 
interactions between adjacent nodes of the chain may be chosen in such a way, 
to ensure localization of excitation only for two ends of chain and perfect 
transfer from first to last node \cite{ek04}.

Let us recollect some essential ideas. The quantum information science is most 
often related with quantum systems with {\em finite} number of (basic) states
and it was quite clear from examples with qubits above. In more general case the
term {\em qudit} is often used for a quantum system with $d$ states, {\em e.g.}, 
a particle with spin $s$ corresponds to $d=2s+1$. Qubit is the particular case 
with $d=2$ and $s=1/2$. 

\begin{figure}[!t]
\centering
a)\includegraphics{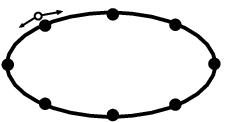}~\hfil~
b)\includegraphics{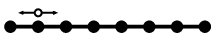}
\caption{Models with lattices}
\label{fig_latt}
\end{figure}

Other model of qudit is some particle in a lattice with $d$ locations. 
Two simple examples with $d$ nodes are {\em ring} (Figure~\ref{fig_latt}a)
\cite[Chap.~15-4]{FLP3} and a {\em line} (Figure~\ref{fig_latt}b) 
\cite[Chap.~15-5]{FLP3}. If we consider a {\em single} 
electron in such a circular or linear system, the wave numbers $k_j$
of stationary states may be expressed as $k_j b = 2\pi j / d $ and 
$k_j b = \pi j / (d + 1) $ respectively \cite[Chap.~15]{FLP3}. The energies 
in both cases are 
\begin{equation}
E_j = E_0 - 2A \cos (k_j b).
\label{Ek}
\end{equation}
Here $b$ is distance between atoms and $A$ is amplitude of transition.

\section{Understanding perfect state transfer}
\label{Sec:perf}

Such chains may be used for quantum communications \cite{bo03},
but a nonlinear dispersion law like Eq.~(\ref{Ek}) may be considered
as a certain obstacle for the good transmission. Already mentioned
earlier {\em perfect} scheme of transport has {\em varying} amplitudes 
of transition \cite{ek04}
\begin{equation}
A_j = A \sqrt{j\,(d-j)}.
\label{Aj}
\end{equation}

Such a model may appear more natural, if to consider a simpler equivalent 
\cite{vl07}. Indeed, in the continuous case the ideal transmission 
of a signal might be obtained with the linear law of dispersion $E(k) \propto k$. 
For the quantum case with the discrete lattice there is similar approach. 

Let us denote a state with occupation of only $l$'th node as $|l\rangle$,
{\em cf\/} \cite[Figure 13-1]{FLP3}.
A spin chain also may be used \cite{ek04}
instead of the lattice with $d$ states. 
Yet, the chain has $2^d$ basic states (Figure \ref{fig_chain}a), only $d$-dimensional 
subspace is used \cite{ek04} and it illustrates rather standard correspondence between 
such lattices \cite[Chap.~13]{FLP3} and spin waves in chains with exchange interaction
\cite[Chap.~15]{FLP3}.

If to consider a {\em ring} (Figure~\ref{fig_latt}a) with $d$ nodes, the
ideal scheme of transfer could be described via 
{\em cyclic shift} operator  
\begin{equation}
{\bf U} : |l\rangle \mapsto |\,l + 1 \mod d\rangle,
\quad l = 0,\ldots,d-1.
\label{shift}
\end{equation}

Eigenvectors and eigenvalues of the operator may be simply found 
\begin{equation}
{\bf U} |\kappa_j\rangle = \zeta_j |\kappa_j\rangle, \quad
\zeta_j = \exp\bigl(\frac{2\pi i}{d}\, j\bigr),
\label{Uk}
\end{equation}
\begin{equation}
|\kappa_j\rangle = 
\frac{1}{\sqrt{d}} \sum_{l=0}^{d-1} \exp\bigl(\frac{2\pi i}{d}\, l j\bigr)\, |l\rangle
\label{mom}
\end{equation}
and produce ``momentum'' basis with $d$ states.  

Similar states were already mentioned above in relation with a ``molecular'' ring, 
Figure \ref{fig_latt}. These states had the fixed wave number $k$. A simplest analogy
of a continuous model with linear dispersion may be provided by Hamiltonian with 
eigenvectors  $|\kappa_j\rangle$ described by Eq.~(\ref{mom}) and equidistant 
eigenvalues, {\em i.e.}, 
\begin{equation}
{\bf H}\, |\kappa_j\rangle = \hbar\theta j \, |\kappa_j\rangle
\label{Hkj}
\end{equation}
with unessential constant $\theta$.

Evolution of a system due to such Hamiltonian in the same basis 
may be expressed via a diagonal matrix, {\em i.e.}, 
\begin{equation}
|\kappa_j\rangle \mapsto \exp(-i \theta j t) |\kappa_j\rangle.
\label{jt}
\end{equation}
It is convenient to choose time step 
\begin{equation}
\Delta t = \frac{2 \pi}{\theta\, d},
\label{Delt}
\end{equation}
because the matrix of evolution of the quantum system for such a time 
period coincides with introduced earlier Eq.~(\ref{shift})
matrix of cyclic shift 
\begin{equation}
{\bf U} = \exp(\frac{{\bf H} \Delta t}{i\hbar}),
\label{Uexp}
\end{equation}
{\em i.e.}, it is expressed in basis $|\kappa_j\rangle$ as
\begin{equation}
{\bf U} : |\kappa_j\rangle \mapsto 
\exp\bigl(\frac{2\pi i}{d}\, j\bigr)\, |\kappa_j\rangle.
\tag{\ref{Uk}$^\prime$}
\end{equation}
So, on the one hand, ${\bf U}$
is a discrete analogue of operator with linear dispersion,
on the other one, it ensures perfect transmission Eq.~(\ref{shift}) of 
local state along chain. 

The advantage of such approach is very clear law of evolution Eq.~(\ref{shift}),
but in the basis $|l\rangle$ there is no simple expression like Eq.~(\ref{Hkj})
for Hamiltonian used in Eq.~(\ref{Uexp}).
The Hamiltonian in such basis may be found \cite[Eq.~(10)]{vl07}, but it
has nonzero values of transition amplitudes for any two sites, unlike initial 
models with nonzero elements only for adjacent locations \cite{FLP3}.

It is useful to look for Hamiltonians
with the same equidistant spectra, but nonzero values of transition only 
for $j \pm 1$. Analogues of angular momentum components operators like 
${\bf J}_x$ or ${\bf J}_y$ have necessary properties and in \cite{ek04} was used 
such a formal Hamiltonian with only nonzero elements $\grave{H}_{j,j\pm 1}$ corresponding 
to Eq.~(\ref{Aj}) and proportional to ${\bf J}_x$ for some fictitious particle 
with spin $s=(d - 1)/2$
\begin{equation}
\setlength{\arraycolsep}{1pt}
\grave{\bf H} =
\vartheta\hbar\!\!
 \begin{pmatrix}
 0 \ \quad \sqrt{d-1} \ \quad 0 &\cdots\! &  0 \ \ \qquad 0 \\
 \sqrt{d-1} \quad 0 \ \sqrt{2 (d-2)}& \cdots\! &  0 \ \ \qquad 0\\
 0 \ \ \sqrt{2 (d-2)} \ \ 0 & \cdots\!  & 0 \ \ \qquad 0 \\
 \dotfill\ddots &\ddots &\ddots\dotfill \\
 0 \ \ \qquad 0 \ \  \qquad 0 \, & \cdots\! & \ \ 0 \ \ \sqrt{d-1} \\
 0 \ \ \qquad 0 \ \  \qquad 0 \, & \cdots\! &  \sqrt{d-1} \ \ 0 \ \ %
 \end{pmatrix}.
\label{Hj}
\end{equation}

Evolution of a system with such a Hamiltonian is described by operator
\begin{equation}
{\bf R}(t) = \exp(\frac{\grave{\bf H} t}{i \hbar}) 
 = \exp(\frac{\vartheta{\bf J}_x t}{i\hbar}).
\label{Rx}
\end{equation}
It coincides with a revolution generated in $d$-dimensional 
representation of rotation group by ${\bf J}_x$ and familiar from 
theory of angular momentum \cite[Chapt.~18]{FLP3}.
Here is discussed a linear chain, but due to such representation there
is an analogy with a ring --- it can be considered formally as
a chain with reflection on the boundaries.

There is a subtle problem, because instead of one-way transmission the state is 
rather circulating between two ends of the lattice with period $\pi/\vartheta$. 
It may be resolved by controlled state transfer, like {\em quantum bots (qubots)} 
discussed in \cite{vl07} or more cumbersome schemes.

The consideration above illustrates application of general 
methods for description of information transfer in the different types of 
{\em quantum wires}. From the one hand it takes into account quite
specific properties of quantum systems, from the other one it
draws a parallel with the traditional information science.

\section{Weyl commutation relations}
\label{Sec:Weyl}

Let us return to the question about problems with quantum uncertainty. 
The operator of shift ${\bf U}$  Eq.~(\ref{shift}) is an important attribute
of quantum mechanics and used in so-called Weyl commutation
relations \cite{WeylGQM}.

For the continuous case such a formalism has the direct correspondence
with the Heisenberg commutation relations \cite{WeylGQM,QFT}.
Indeed in some cases it is necessary to use the Heisenberg uncertainty relation 
with proper care due to subtleties with definition of domains of
operators \cite{QFT,PerQT}, {\em e.g.}, for a ring with the periodic coordinate
$0 \le q < 2 \pi$ the value $\Delta q$ is always finite, but $\Delta p = 0$
for eigenfunctions of ${\bf p}$, {\em i.e.},
\begin{equation}
 u_k(q) = (2 \pi)^{-1/2} \exp(i k q), 
\label{ukq}
\end{equation}
and so $\Delta p \,\Delta q = 0$ \cite{QFT,PerQT}.
Here $|u_k(q)|^2$ is constant and $\Delta q = \pi/\sqrt{3}$ 
--- it coincides with the standard deviation of the random variable 
$0 \le q < 2\pi$ with the uniform distribution.

More details may be found in \cite[Chapt.~4-3]{PerQT} (with
notation $\theta$ and $p_\theta$ for ${\bf q}$ and ${\bf p}$ respectively).
It should be mentioned, that operators ${\bf q}$ and ${\bf p}$ above 
were not ``usual'' coordinate and momentum for a particle {\em on a line}
with uncertainty relation $\Delta p \,\Delta q \ge \hbar/2$.
For a line $\Delta q$ is not limited by some fixed value and
for $\Delta p \to 0$ we would have ``unrestricted'' plane 
waves instead of Eq.~(\ref{ukq}), {\em i.e.}, $\Delta q \to \infty$.
Here the limit is an uniform distribution on an infinite line 
$-\infty < q < \infty$, instead of a bounded ring.

Qubit and qudit are said to be ``discrete quantum variables'' 
widely used in the quantum information science.
For such systems the problem with proper definition 
of coordinate, momentum operators and analogues of uncertainty
relations could look even worst, than for continuous ring,
but {\em Weyl quantization} \cite{WeylGQM} may help to resolve 
that. The idea is to consider exponents of coordinate 
and momentum operators \cite{WeylGQM,QFT}
\begin{equation}
{\bf U}(a) = \exp(i a {\bf p}), \quad
{\bf V}(b) = \exp(i b {\bf q}).
\label{UV}
\end{equation}

In the continuous case the actions of such operators on a  wave
function $\psi(x)$ are represented as 
\begin{equation}
 {\bf U}(a) \psi(x) = \psi(x+\hbar a),\quad
 {\bf V}(b) \psi(x) = e^{i b x} \psi(x),
\label{UVpsi}
\end{equation}
and direct consequence of Heisenberg commutation relation
\begin{equation}
  [{\bf q},{\bf p}] \equiv
  {\bf q} {\bf p} - {\bf p} {\bf q} = i \hbar
\label{xpcom}
\end{equation}
is Weyl commutation relation \cite{WeylGQM,QFT}
\begin{equation}
 {\bf U}(a) {\bf V}(b) = \exp(i \hbar a b) {\bf V}(b) {\bf U}(a).
\label{UVcom}
\end{equation}

Such a scheme has some advantage because {\em Weyl pair} ${\bf U}$, ${\bf V}$ 
with appropriate properties may be formally written even if ${\bf p}$ and 
${\bf q}$ are not (well) defined. It may be even used for discrete quantum 
variables \cite{WeylGQM} and it was already reproduced above in example 
with operator of shift {\bf U} Eq.~(\ref{shift}). The second operator
in this case is 
\begin{equation}
{\bf V} : |l\rangle \mapsto 
\exp\bigl(\frac{2\pi i}{d}\, l\bigr)\, |l\rangle.
\label{clock}
\end{equation}
So, ${\bf U}$ and ${\bf V}$ are $d{\times}d$ matrixes with
commutation relation
\begin{equation}
 {\bf U} {\bf V} = \exp\bigl(\frac{2\pi i}{d}\bigr)\, {\bf V} {\bf U}.
\label{UVVU}
\end{equation}

These matrixes together with relation Eq.~(\ref{UVVU}) were introduced 
by Weyl \cite{WeylGQM}. Really, the ``shift''  ${\bf U}$ and ``clock'' 
${\bf V}$ matrixes were considered even earlier in few works of J. J. Sylvester 
around 1882--1884. In quantum information science they are also 
known as ``generalized Pauli matrixes'' with an alternative
notation ${\bf X}$ and ${\bf Z}$ \cite{high}. 

Other examples may be found elsewhere \cite{low}. Let us only consider 
less formally questions about uncertainty for discrete quantum variables. 
Famous Stern-Gerlach experiment demonstrates only two possible projections 
on some axis for spin one-half. For spin $s$ there are $2s+1$ projections. 
It is just obvious statement about {\em quantization of angular momentum}.

A belief about inevitable problems with quantum transport due to uncertainties 
of trajectories related with lack of examples with similar effects for some 
spatial properties of quantum systems. It is more common to expect quantization 
for energy levels, angular momentum, {\em etc.} 
Yet, in quantum information science were quite natural formal models with 
discrete spatial variables, {\em e.g.}, quantum cellular automata, quantum 
lattice gases \cite{RW95,Mey96}, {\em etc.}

\section{Conclusion}

The quantum information technologies may be useful
for construction of difficult nano-technology devices, because they are providing 
universal and compact way of understanding different processes with ``systems of
quantum systems.'' 
It is an analogy with the application of usual information technologies for description 
in symbolic form of classical processes and objects.

In the paper were recollected few simple models: the representation of data by qubits 
and signal transfer in small quantum systems. These models may be quite familiar 
in area of quantum computing, but it should be emphasized, that main purpose of 
present work --- is {\em not} theory of quantum algorithms adapted for 
cryptography. Even usual electronic computers were initially constructed for 
code-breaking and plain calculations, but nowadays they work as well in absolutely 
different areas. 

It should be mentioned also, that this presentation is {\em not} concentrated on
restricted question, how nanotechnologies could help to build a quantum computer
to crack some ciphers. 
It is  rather analyzed, how ``quantum-computer-type of thinking'' may help to understand 
and control nano-scale systems and devices.

\end{document}